# New VSC-HVDC interconnection between the Iberian Peninsula and Balearic Archipelago to enable energy transition

**Javier RENEDO***
**Red Eléctrica - Redeia**
**Spain**
**javier.renedo@ree.es**

**Silvia SANZ VERDUGO**
**Red Eléctrica – Redeia**
**Spain**
**sisanz@ree.es**

**Antonio CORDÓN**
**Red Eléctrica – Redeia**
**Spain**
**acordon@ree.es**

**Belén SEGURA**
**Red Eléctrica – Redeia**
**Spain**
**bsegura@ree.es**

**David CASTAÑEDA**
**Red Eléctrica – Redeia**
**Spain**
**david.castaneda@ree.es**

**Rosalia RIVAS**
**Red Eléctrica – Redeia**
**Spain**
**rrivas@ree.es**

**Patricia LABRA**
**Red Eléctrica – Redeia**
**Spain**
**plabra@ree.es**

**SUMMARY**

One of the challenges of the Spanish Transmission System Operator (TSO) is the decarbonisation of the Balearic Archipelago, by means of the integration of Renewable Energy Sources (RES) in the islands, as well as increasing the transmission capacity between the Iberian Peninsula and the Balearic Islands. Since the Balearic Archipelago is an island power system, the decarbonisation brings challenges related to power system stability and operation.

A new High Voltage Direct Current (HVDC) interconnection between the Iberian Peninsula power system and the Balearic Islands power system is planned to facilitate the decarbonisation of the Balearic Archipelago (PEN-BAL2 Project). The HVDC link will be based on Voltage Source Converter (VSC) technology and will consist of a bipole of 2x200 MW, a DC voltage of ±250 kVdc and +100/-150 Mvar of reactive power capacity for each converter station. One converter station will be connected to a future El Fadrell 400 kV substation (Castellón, Valencia, Iberian Peninsula), while the other converter station will be connected to the existing San Martín 220 kV substation (Mallorca Island, Balearic Islands). The link will have HVDC submarine cables of 363 km (approx.).This paper will describe the challenges for energy transition in the Balearic Archipelago, technology enablers in general and PENBAL2 VSC-HVDC interconnection.

This is a preprint of the following paper:

- J. Renedo, S. Sanz Verdugo, A. Cordón, B. Segura, D. Castañeda, R. Rivas, P. Labra, "New VSC-HVDC interconnection between the Iberian Peninsula and Balearic Archipelago to enable energy transition," *Proc. CIGRE Paris Session, B4-PS3, paper 10623*, Paris, France, 25-30 Aug. 2024, pp. 1-12.

## 1. INTRODUCTION

One of the challenges of the Spanish Transmission System Operator (TSO) is the decarbonisation of the Balearic Archipelago, by means of the integration of Renewable Energy Sources (RES) in the islands, as well as increasing the transmission capacity between the Iberian Peninsula and the Balearic Islands. Fig. 1 shows a map of the Iberian Peninsula and the Balearic Archipelago. Balearic Archipelago consist of four main islands: Mallorca, Menorca, Ibiza and Formentera which are interconnected through various AC submarine cables, forming a synchronous power system. Balearic power system will be referred in document as *SEB* (Sistema Eléctrico Balear in Spanish). Currently, the Technology enablers High Voltage Direct Current link based on Line-Commutated Converters (LCC-HVDC) [1], as shown in Fig. 1. The LCC-HVDC link is connected to Morvedre 400 kV substation (Valencia, Iberian Peninsula) and to Santa Ponsa 220 kV substation (Mallorca Island, Balearic Archipelago).

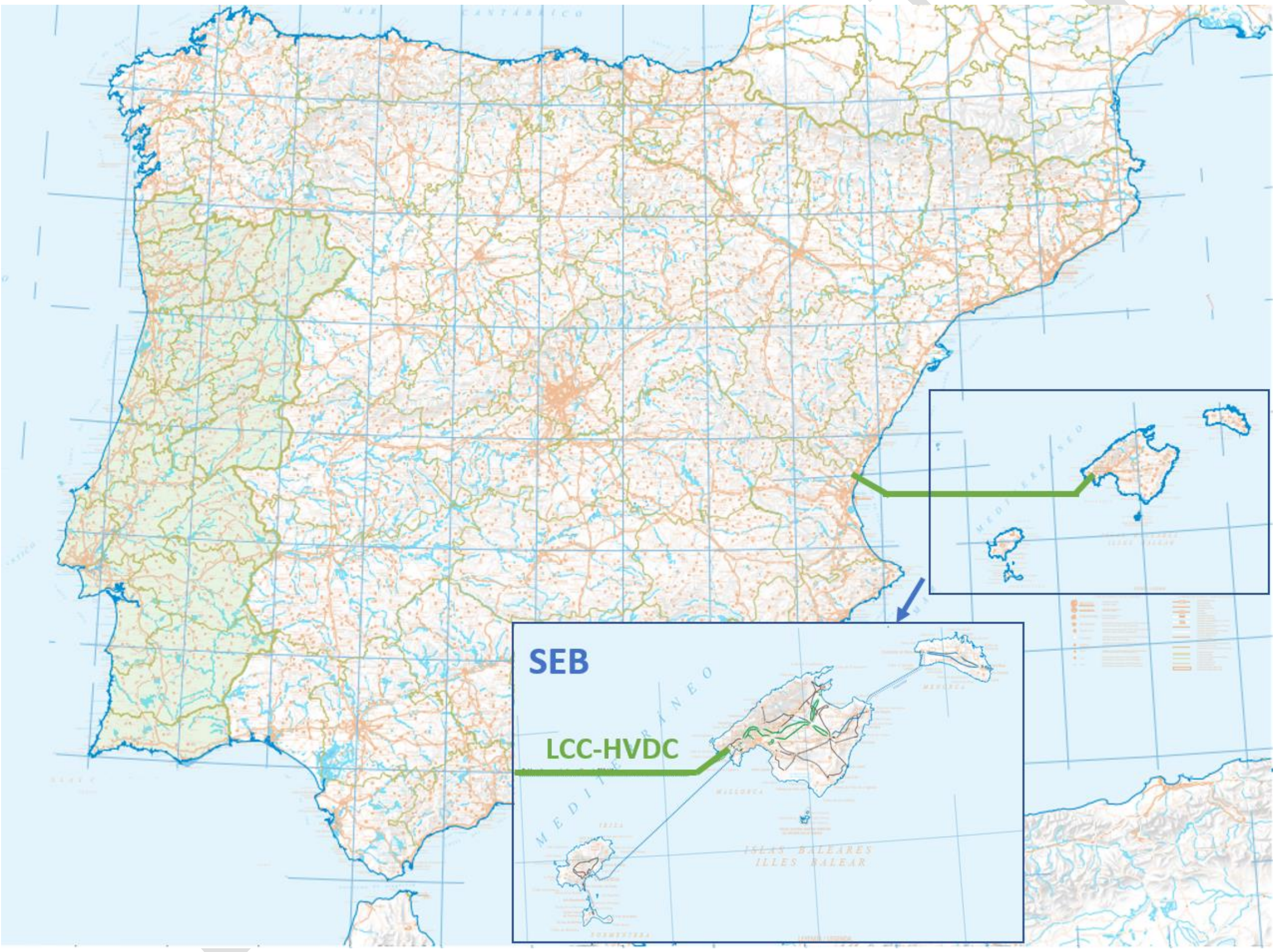

*Fig. 1: Iberian Peninsula and Balearic Archipelago (Spain).*

The peak demand of the Balearic Archipelago power system estimated for Horizon H2026 is around 1271 MW, while the peak demand of the Iberian Peninsula power system, estimated for H2026, is around 46800 MW. Fig. 2 shows the estimated demand of the SEB for H2026 [2]: the estimated peak demand is 1271 MW and the estimated energy consumption during the year is 6253 GW-h. The highest demand in the SEB occurs in summer, due to tourism. As an example, Fig. 3 shows the daily demand curve of a summer day (5-8-2023). Maximum demand of the day occurs at 9:20 pm, where the demand of the SEB is 1008 MW. Fig. 4 shows the generation mix in the SEB of 5-8-2023 at 2:00 pm, which also has a high demand. In this scenario, the dominant generation technology is Combined Cycle Gas Turbine (CCGT) (30.08 %) and the power transmitted from Peninsula to Balearic Islands through the existing LCC-HVDC interconnection. It should be noticed that PhotoVoltaic (PV) generation is also significant (18.26 %). The decarbonisation of the Balearic Archipelago will mainly consist of (a) transmission of RES from the Iberian Peninsula to the Balearic Islands and (b) integration of PV generation in Balearic Islands.

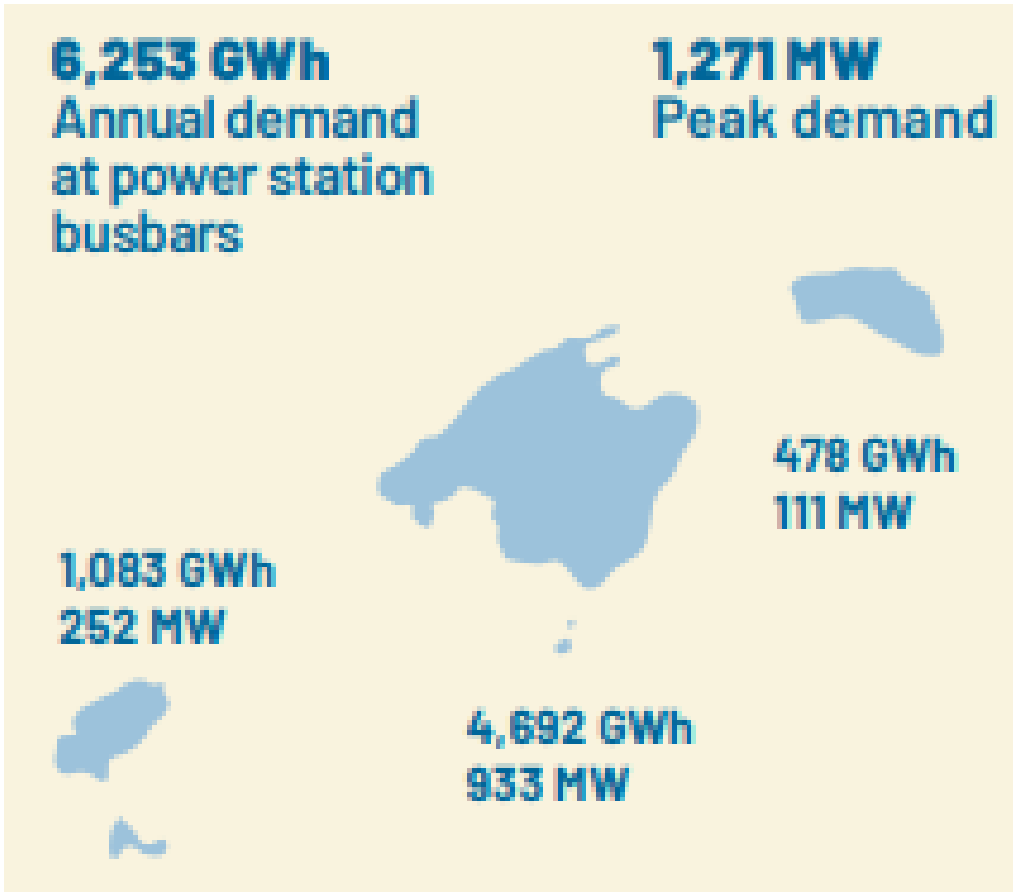

*Fig. 2 : SEB – Estimated demand H2026.*

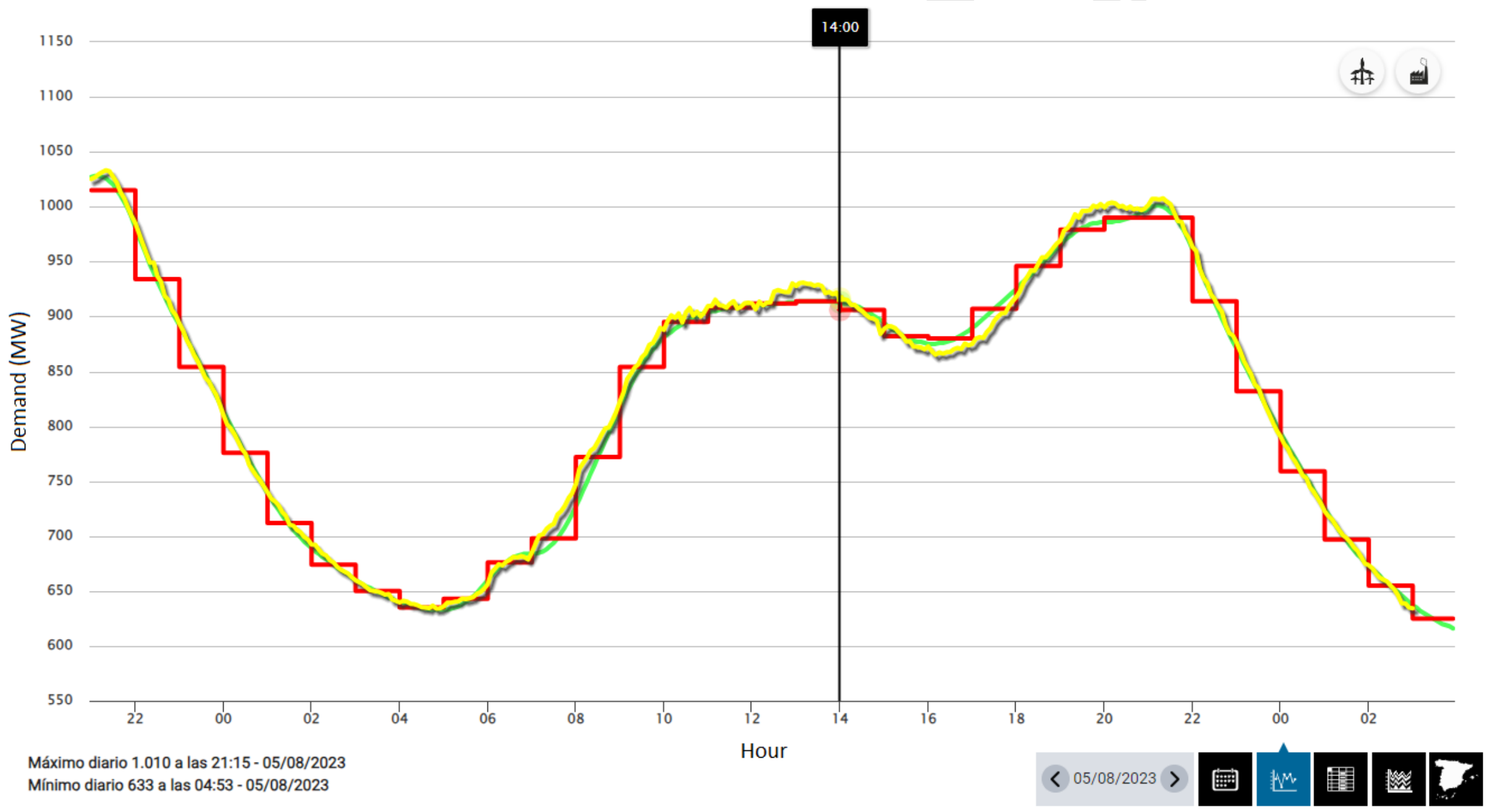

*Fig. 3 : SEB – Demand curve of 5-8-2023 (www.ree.es).*

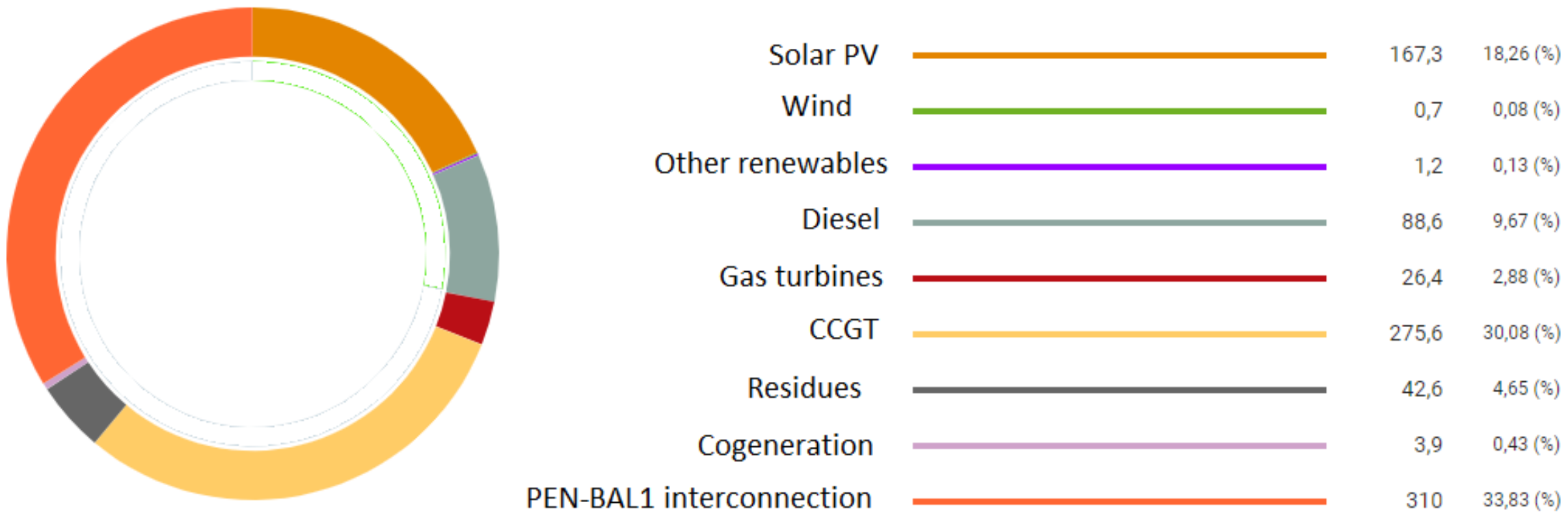

*Fig. 4 : SEB – Generation mix of 5-8-2023 at 2 :00 pm. Total demand: 916 MW (www.ree.es).*

In general, island power systems are small in comparison with interconnected power systems and they are more vulnerable to some stability phenomena, such as frequency stability, for example, and the integration of renewable energy based on power converters (non-synchronous generation) is a challenge [3]. SEB system is an isolated power system with low inertia and, therefore, frequency stability together with the integration of renewable energy is a challenge. Currently, Balearic power system is operated with a minimum number of synchronous generators connected (Must-Run units), which guarantee the required inertia of the system and frequency control to ensure frequency stability, and also voltage control in the SEB. To make it possible energy transition in the Balearic Archipelago, power system reinforcement with technology enablers is needed. Technology enablers will increase (a) the transmission capacity between Peninsula and the Balearic Archipelago, (b) the transmission capacity between islands of the SEB and (c) improve power system stability. Technology enablers in the SEB have been included in the Planning of the Spanish power system for H2026 [2]. Real-world experiences of operation of isolated power systems with large amounts of RES and the use of technology enablers for secure operation of the system can be found in [4].

This paper will describe the challenges for energy transition in the Balearic Archipelago, technology enablers in general and PENBAL2 VSC-HVDC interconnection.

A description of technology enablers for the decarbonisation of Balearic electrical system was presented by the authors in the Regional Ibero-American Meeting of CIGRE 2024 (XIX ERIAC-2024) [5]. This paper also deals with energy transition in the Balearic Archipelago, but it focuses on PENBAL2 VSC-HVDC interconnection project.

## 2. BALEARIC POWER SYSTEM WITH 100% OF RENEWABLE ENERGY AND OPERATIONAL CHALLENGES

In a decarbonised SEB, the system will have large amounts of non-synchronous generation and it will be possible to achieve 100 % non-synchronous generation in certain scenarios. This has three main operational challenges:

- Frequency control.
- Minimum level of short-circuit power (Scc) needed:
  - Need of a grid with a certain strength (to guarantee stability, secure operation of the existing LCC-HVDC link, etc…).
  - Minimum level of short-circuit currents to ensure the correct behaviour of the protection system.
- Voltage control.

### 2.1. Frequency control

Frequency stability is defined as the ability of the power system to reach an stable equilibrium point and maintain the frequency within admissible limits when unbalances between generation and demand occur [6], [7], [8], [3]. Key Performance Indicators (KPI) for frequency stability are the maximum frequency deviation during the transient, Rate Of Change Of Frequency (ROCOF) and frequency deviation in steady state. The main factors that affect frequency stability of a power system are total inertia, total primary frequency control and available reserve, load-shedding schemes and protection systems of under/over-frequency of the devices connected to the grid. It is well known that frequency stability is much more critical in insular power systems, due to their small size and the integration of non-synchronous generation introduce additional challenges in this type of systems [3].

Currently, Balearic power system is operated with a minimum number of synchronous generators connected which guarantee the required inertia of the system and frequency control to ensure frequency stability.

### 2.2. Minimum level of short-circuit power

A minimum level of short-circuit power is needed to ensure the stability of SEB system due to two main aspects:

- Enough grid strength to guarantee the stability of the SEB system and the correct behaviour of the existing LCC-HVDC link.
- Short-circuit-current level needed to ensure the correct behaviour of the protection system.

Grid strength at a certain bus is often characterised by its short-circuit power (Scc) or its Short Circuit Ratio (SCR). These indicators provide information about the series impedance of the Thevenin equivalent seen by the bus. A strong grid has high values of Scc and SCR and a weak grid has low values of Scc and SCR.

In Santa Ponsa 220 kV substation, a minimum level of short-circuit power is needed to ensure the correct behaviour of the existing 2x200 MW LCC-HVDC link. This is due to the fact that LCC-HVDC need a grid with a certain strength to be connected to, to ensure successful commutation of thyristors and to guarantee the stability [6], [9].

The short-circuit level at Santa Ponsa 220 kV substation is linked to:

- Grid topology.
- Synchronous generators in service.

In order to avoid stability issues in scenarios with low short-circuit power at Santa Ponsa 220 kV substation, a monitoring system of critical contingencies and for operation rules for the LCC-HVDC link was implemented (so-called ASPAS system [10]). The main operation rules of the existing LCC-HVDC can be summarised as follows.

- Scc≥1300 MVA: Normal operation. The LCC-HVDC link can be operated at full capacity (2x200 MW).
- 500 MVA≤Scc<1300 MVA: Maximum power transferred through the LCC-HVDC link is limited to 40 % of its nominal power (160 MW), and monopolar operation.
- Scc<500 MVA: Tripping of the LCC-HVDC link.

### 2.3. Voltage control

In all power systems, bus voltages must be maintained within operating limits. In SEB system, voltage control is carried out by synchronous generators in service and shunt compensators (connection/disconnection of shunt reactances and capacitor banks).

The operation of an island electrical system with a reduced number of synchronous generators implies an additional challenge for voltage control, being especially relevant in scenarios of low demand (valley) and high capacitive tendency of the network.

Additionally, in the electrical system of the Balearic Islands, the operation of the existing LCC-HVDC link implies important sudden voltage variations in its area of influence (Santa Ponsa 220 kV) as a consequence of the successive connection/disconnection maneuvers of the large capacitor filters. (4x41 Mvar) installed to compensate the reactive consumption of the converters and to perform the necessary harmonic filtering.

Given the expected reduction of synchronous generation in a potential decarbonization scenario in the Balearic Islands, technological reinforcements (synchronous compensators, etc.) with sufficient capacities will be necessary to assume this continuous voltage control in a more flexible and robust way..

## 3. TECHNOLOGY ENABLERS FOR ENERGY TRANSITION IN THE BALEARIC ARCHIPELAGO

Energy transition in Balearic Archipelago will reduce the use of thermal generation in the SEB, reducing, therefore, CO2 emissions and also reducing the operating costs of the system. This will allow the increase of transmission capacity from the Iberian Peninsula to Balearic Islands, transferring additional power from renewable energy sources from Peninsula, as well as increasing the renewable energy in Baleares. Power system reinforcement consisting of technology enablers is necessary to achieve energy transition in the Balearic Archipelago, ensuring the reliability and security of the SEB, and they have been described in the approved planning of the Spanish electrical system in Horizon H2026 [2]. The feasibility of technology enablers for the decarbonisation of the SEB has been proved by means of a Cost-Benefit Analysis (CBA), following the European guidelines and methodology, as described in [2]. The analysis resulted in positive CBA with the proposed technology enablers, in comparison with the case in which they were not considered.

Technology enablers for energy transition in the Balearic Archipelago are shown in Fig. 5 and they consist in:

- New submarine 2x200 MW VSC-HVDC bipolar link with metallic return between the Iberian Peninsula and Balearic Archipelago (PENBAL2), between Mallorca (Island, Balearic Archipelago) and the Iberian Peninsula.
  It will allow to increase the transmission capacity between the Iberian Peninsula and the Balearic Archipelago and to ensure the stability of the SEB.
- Synchronous Condensers (CS) in Mallorca Island (5x100 MVA) :
  - SC Santa Ponsa 220 kV substation: 2x100 MVA.
  - SC Llubí 220 kV substation: 2x100 MVA.
  - SC Valldurgent 220 kV substation: 1x100 MVA.

  They will contribute to system inertia, grid strength and voltage control.
- Battery Energy Storage Systems (BESS) :
  - BESS San Antonio 66 kV substation: 90 MW/67.5 MW-h (Ibiza Island)
  - BESS Mercadal 132 kV substation: 50 MW/37.5 MW-h (Menorca Island)

  BESS systems are defined as fully-integrated network components, according to Article 63 of European Union (UE) Directive 2019/944 [11]. They are planned to be in standby in normal operation and they will inject active power only in case of tripping of interconnection links between the Iberian Peninsula and the Balearic Archipelago or in case of tripping of the AC interconnections between the islands of the SEB. This results in an increase of power transmission between the Iberian Peninsula and the SEB and between the islands of the SEB.

This paper will focus on VSC-HVDC PENBAL2 Project.

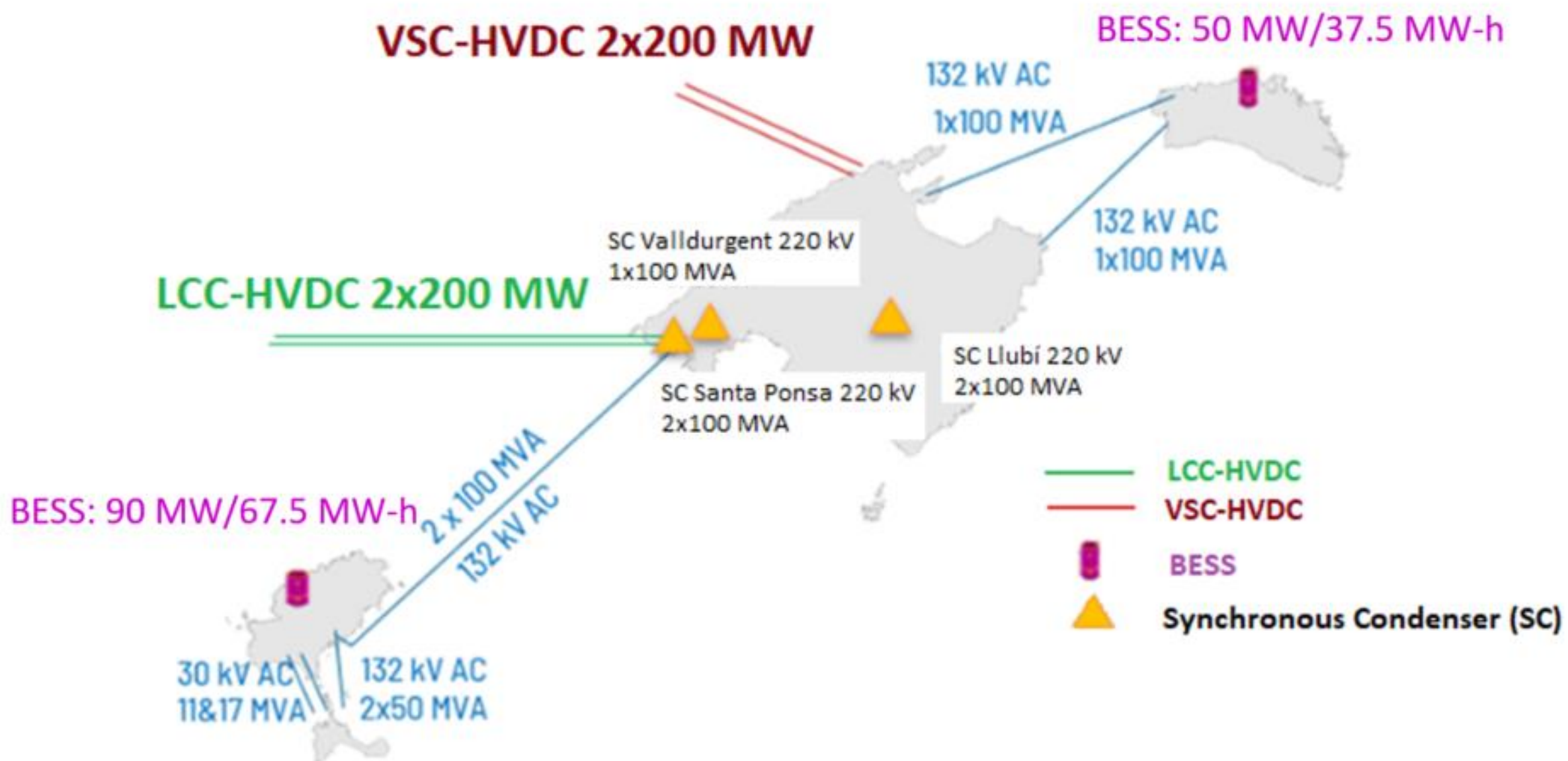

*Fig. 5 :SEB – Technology enablers for energy transition in the Balearic Archipelago .*

## 4. PENBAL2 VSC-HVDC INTERCONNECTION

The new VSC-HVDC link between the Iberian Peninsula and the Balearic Archipelago will consist of a 2x200 MW bipolar link with metallic return and +-250 kV of DC voltage (Fig. 5). The converter station at the Iberian Peninsula will be connected to El Fadrell 400 kV substation (Castellón, Valencia, Peninsula) and the converter station at Balearic Archipelago will be connected to the existing Sant Martín 220 kV substation, the North of Mallorca Island (Balearic Archipelago). The link will have HVDC submarine cables of 363 km (approx.). Fig. 6 shows the scheme of PENBAL2 VSC-HVDC link.

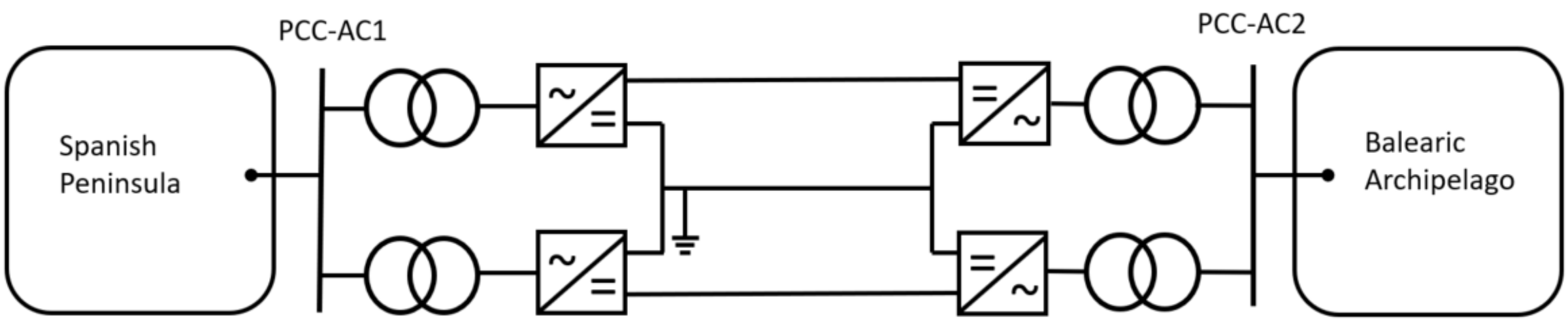

*Fig. 6 : PENBAL2 2x200 MW system.*

The potential of the new VSC-HVDC as a technology enabler for energy transition is linked to its power capacity and to its control system, based on most recent technologies.

Voltage source converters (VSC) can be controlled in two ways:

- Grid-following VSC (GFL) (Fig. 7-a): The converter behaves as a current source and it needs a grid with enough stiffness to be connected to.
- Grid-forming VSC (GFM) (Fig. 7-b): The converter behaves as a voltage source behind a series impedance and it is capable of creating a grid itself (imposing the frequency). The GFM-VSC needs to have a self-synchronisation control algorithm, in order to be able to operate in synchronism with the rest of the grid, independently if the grid has other generators connected. This concept is called *synchronous grid-forming* by some authors [12]. There are different self-synchronisation strategies for GFM-VSCs [13], [14], [15], [16], [17], [18], [19], but all of them with similar philosophy: to mimic the behaviour of conventional synchronous machines, in particular their capability of synchronisation in a natural way. In addition, manufacturers of power converters could propose novel variants during the following years.

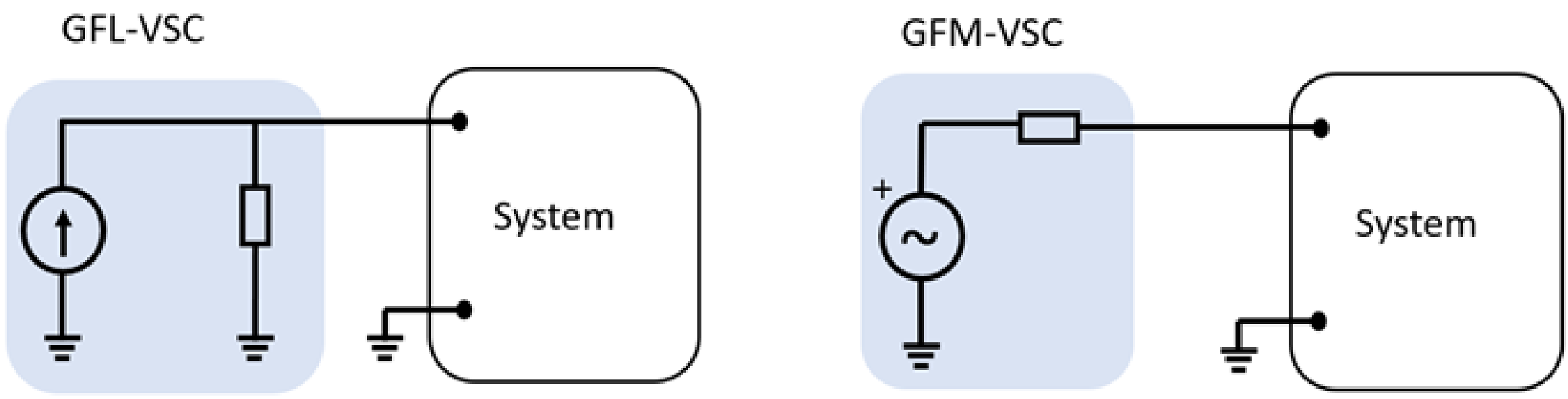

**(a)** **(b)**

*Fig. 7 : Scheme of (a) GFL-VSCs and (b) GFM-VSCs.*

In the scenario of a decarbonised Balearic Archipelago, the system will have a low short-circuit power and a low inertia, aiming to achieve 100 % non-synchronous generation in certain scenarios. In addition, the presence of the existing LCC-HVDC link in such scenarios is a relevant challenge. LCC-HVDC links need a grid a certain stiffness to be connected to. The stiffness of the grid is quantified often with the short-circuit power or the short-circuit ratio (SCR). LCC-HVDC links can have problems associated to commutation failures and stability when they are connected to grids with low SCR. To tackle these problems, it is essential for the new VSC-HVDC link to have grid-forming control in Mallorca's converter station (Fig. 6). Previous studies have analysed the advantages of grid-forming control in PENBAL2 VSC-HVDC link [20].

The main objectives of the new PEN-BAL2 VSC-HVDC interconnection are as follows:

- Increase of power transmission from the Iberian Peninsula to the Balearic Archipelago, aiming to achieve the decarbonisation of the latter.
- Contribution to grid strength in the Balearic electrical system by means of grid-forming control at Mallorca converter station, in order to guarantee power system stability in scenarios with large amounts of non-synchronous generation, even in scenarios with 100 % non-synchronous generation, and with the presence of the existing LCC-HVDC link.
- Contribution to frequency stability in the Balearic electrical system, by means of frequency control and synthetic inertia in Mallorca converter station. With a frequency control in PENBAL2 VSC-HVDC link, frequency support provided from Peninsula to the Balearic power system will be increased.
- Contribution to voltage control to Balearic electrical system.
- In general, to contribute to the reliability and security of the Balearic electrical system by means of the control flexibility of VSC-HVDC systems (active power control, reactive power control, GFM control, emergency controllers, black start, etc…).

## 5. TECHNICAL STUDIES

All technology enablers described in Section 3, including PEN-BAL2 VSC-HVDC Project, have the objective to achieve the decarbonisation of the Balearic Archipelago. Nevertheless, they are not free of challenges related to the stability of an isolated power system with large amounts of power-electronic devices. Since power converters present a high control flexibility, PENBAL2 VSC-HVDC link must be designed to guarantee the stability of the system.

PENBAL2 VSC-HVDC and the rest of technology enablers must guarantee:

- The stability of the SEB system in scenarios with low inertia.
- The stability of the SEB system in scenarios with low short circuit power (weak grid).
- The stability of the SEB system in scenarios with 100 % non-synchronous generation.
- To avoid stability issues in SEB system related to the interaction of the new VSC-HVDC link with the existing LCC-HVDC link and with other power-electronic devices.
- To avoid stability issues in Peninsula side related to the interaction of the new VSC-HVDC link with the existing LCC-HVDC link and with other power-electronic devices.
- Flexibility in the control of the VSC-HVDC link to guarantee the stability of the system in scenarios with and without non-synchronous generation because there will be scenarios of different nature.

To tackle these challenges, the Spanish TSO is carrying out the following actions:

- Frequency stability studies.
- Transient stability studies (Root-Mean-Square (RSM) simulation).
- ElectroMagnetic Transient Simulation (EMT) studies.
- Studies related to the control schemes of the new VSC-HVDC link.
- Development of technical specifications of the technology enablers to guarantee the reliability and security of the power system.
- Contact with the technologists of power converters to know the state of the art and development in the control systems of power converters.

## 6. CONCLUSIONS

A new 2x200 MW VSC-HVDC interconnection between the Iberian Peninsula power system and the Balearic Islands power system is planned to facilitate the decarbonisation of the Balearic Archipelago (PEN-BAL2 Project). The new VSC-HVDC will consist of a 2x200 MW bipole with metallic return.

The new VSC-HVDC interconnection is planned to increase power transmission between the Iberian Peninsula and Balearic Archipelago and to guarantee the reliable and secure operation of the system.

Specifically, the main objectives of PEN-BAL2 VSC-HVDC project can be summarised as follows:

- Increase of power transmission between the Iberian Peninsula and Balearic Archipelago.
- Contribution to the strength of Balearic electrical system by means of grid-forming control in Mallorca converter station.
- Contribution to frequency control (from Peninsula to Balearic power system).
- Contribution to voltage control.